# Равновесная функция распределения кластеров по размерам в конечной системе

© **Умирзаков Ихтиёр Холмаматович**

*Лаборатория моделирования. Институт теплофизики СО РАН. Пр. Лаврентьева, 1.*
*г. Новосибирск, 630090. Россия. Тел.: (383) 354-20-17. E-mail: tepliza@academ.org*

**Аннотация**

Методами статистической термодинамики найдена равновесная функция распределения кластеров (наночастиц) по размерам в системе, состоящей из конечного числа молекул (атомов), находящихся в конечном замкнутом объеме при постоянной полной энергии (изолированная система). На основе кинетического уравнения нуклеации используя однокапельное приближение функция распределения найдена как для изолированной, так и для изотермической системы. Полученные результаты сравниваются с данными компьютерного моделирования конечных двухмерных систем.

**Введение**

Равновесное распределение кластеров по размерам в конечной системе ранее изучалось в работах [1-15]. В основном изучалась изотермическая система $(T, V, N) = const$ (здесь $N$ – число атомов (молекул), $T$ – температура, $V$ – объем, занимаемый системой).

Поведение функции распределения кластеров по размерам, полученной в [3, 4, 7] качественно согласуется с данными численного моделирования [2, 4, 7].

Однако количественно наблюдаются большие расхождения. Причина этих разногласий была установлена в [11, 12, 14]. Здесь же было введено и обосновано "однокапельное" приближение, в рамках которого были устранены указанные выше разногласия. Отметим, что в работах [3, 4, 7, 11, 12, 14, 15] использовалось "капиллярное приближение".

Работа состоит из трёх частей. В первой части приведены результаты исследования методами статистической термодинамики равновесного распределения кластеров по размерам в изолированной системе $(E, V, N) = const$ ($E$ – полная энергия системы), соответствующей микроканоническому ансамблю.

Вторая часть работы посвящена изучению равновесного распределения как в изотермической, так и в изолированной системах на основе кинетического уравнения нуклеации Френкеля-Зельдовича [19, 20], а третья часть содержит сравнение полученных результатов с данными численного моделирования работ [5, 9, 10].

## Результаты и их обсуждение
### 1. Изолированная система

Для определения равновесного распределения кластеров в изолированной системе будем считать, что система представляет собой смесь идеальных газов кластеров различных размеров. Тем самым мы считаем, что система находится вдалеке от критической точки. Тогда статистический вес (фазовый объём) системы можно записать в виде ($\vec{a} = (E,V,N)$)

$$\Omega(\vec{a}) = \sum_{\vec{N}} \prod_{i=1}^{N} \prod_{k_i=1}^{\infty} G_i^{N_{k_i}}(E_{k_i}) / N_{k_i}!, \qquad (1)$$

где $N_{k_i}$ – число кластеров, содержащих $i$ молекул, каждый из которых имеет полную энергию $E_{k_i}$, $G_i(E_{k_i})$ – статистический вес одного кластера размера $i$ с энергией $E_{k_i}$.

Выше учтено, что каждый кластер размера $i$ может принимать только дискретный набор значений энергии $E_{k_i}$, $k_i = 1,2,\ldots$, поскольку он находится в замкнутом сосуде с конечным объемом.

Суммирование производится по всем наборам кластеров $\vec{N} = \{N_{k_i}, i = 1,..,N; k_i = 1,2,...\}$, удовлетворяющим условиям сохранения полного числа частиц в системе $N$ и полной энергии системы $E$

$$\sum_{i=1}^{N} i \sum_{k_i=1}^{\infty} N_{k_i} = N, \tag{2}$$

$$\sum_{i=1}^{N} \sum_{k_i=1}^{\infty} E_{k_i} N_{k_i} = E. \tag{3}$$

Суммирование в (1) осложнено по сравнению со случаем изотермической системы в силу наличия еще одного условия (3).

В дальнейшем в (1) будем учитывать только слагаемое, соответствующее наиболее вероятному набору энергий $\tilde{E} = \{\tilde{E}_{k_i}, i = 1,..,N; k_i = 1,2,...\}$ для данного набора кластеров $\vec{N}$, считая вклад остальных слагаемых малым по сравнению с наиболее вероятным. Применяя к (1) метод неопределенных множителей Лагранжа (то есть, варьируя отдельный член в (1) по $E_{k_i}$ с учетом условия (3)) получим

$$\partial \ln[G_i(\tilde{E}_{k_i})^{N_{k_i}} / N_{k_i}!] / \partial \tilde{E}_{k_i} - \beta(E, \vec{N}) N_{k_i} = 0, \tag{4}$$

где множитель Лагранжа $\beta$ один и тот же для любых $i$ и $k_i$. Из этой формулы легко получить соотношение

$$\partial \ln G_i(\tilde{E}_{k_i}) / \partial \tilde{E}_{k_i} = \beta(E, \vec{N}). \tag{5}$$

Далее считая, что функциональная зависимость $G_i(\tilde{E}_{k_i})$ от $\tilde{E}_{k_i}$ одна и та же для любого уровня энергии $k_i$ для данного размера $i$ и с учетом последней формулы приходим к выводу, что все кластеры $i$-того размера находятся на одном и том же энергетическом уровне $E_i = \tilde{E}_{k_i}$, кроме случая маловероятной экспоненциальной зависимости $G_i(\tilde{E}_{k_i})$ от $\tilde{E}_{k_i}$. С учетом последнего обстоятельства формулы (1)-(5) можно написать в виде

$$\Omega(\vec{a}) = \sum_{\vec{N}} \prod_{i=1}^{N} G_i(E_i)^{N_i} / N_i!, \tag{6}$$

$$\sum_{i=1}^{N} i N_i = N, \tag{7}$$

$$\sum_{i=1}^{N} E_i N_i = E, \tag{8}$$

$$\partial \ln[G_i(E_i)^{N_i} / N_i!] / \partial E_i - \beta(E, \vec{N}) N_i = 0, \tag{9}$$

$$\partial \ln G_i(E_i) / \partial E_i = \beta(E, \vec{N}), \tag{10}$$

где $N_i$ – число кластеров i-го размера. Теперь суммирование в (6) ведется по всем наборам кластеров $\vec{N} = \{N_1,...,N_N\}$, удовлетворяющим условиям (7), (8), а значение энергии $E_i$ в аргументе $G_i(E_i)$ в (6) удовлетворяет уравнению (10), в котором $\beta$ определяется из условия (8) с учетом (10).

В дальнейшем будем учитывать только те состояния системы, для которых набор кластеров $\vec{N}$ мало отличается от стабильного и/или метастабильного равновесных наборов, то есть будем рассматривать только те состояния системы, которые представляют собой малые отклонения от стабильного или метастабильного равновесных состояний, причем считаем, что система может достаточно долго находиться в состоянии метастабильного равновесия. Для таких состояний можно ввести понятие температуры [16]. Тогда учитывая, что энтропия газа кластеров размера $i$

$$S_i(E_i, N_i) = k \ln[G_i(E_i)^{N_i} / N_i!] \qquad (11)$$

и формулу $(\partial S / \partial E)_V = 1/T$, из (9) заключаем, что $\beta = 1/kT$, где $k$ – постоянная Больцмана.

Для энергии кластера размера $i$ примем формулу [11]

$$E_i = \frac{3}{2}kT \cdot i - q_\infty(i-1) + \alpha(1 - Td\sigma_{l\infty}/dT/\sigma_{l\infty})(i^{2/3} - 1), \qquad (12)$$

представляющую собой интерполяцию между двумя известными значениями $E_i$: при $i = 1$ и $i \to \infty$ (классическое "капиллярное" приближение). Здесь $q_\infty$ – энергия связи молекулы с блоком жидкости, $\sigma_{l\infty}$ – коэффициент поверхностного натяжения для плоской поверхности жидкости, $\alpha = (36\pi v_{l\infty}^2)^{1/3}\sigma_{l\infty}$, где $v_{l\infty}$ – объём на одну частицу в блоке жидкости.

Поскольку $\beta$ определяется из (8), то и температура тоже определяется из него, причём её значение зависит от набора $\vec{N}$, то есть $T = T(E, \vec{N})$. Таким образом, температура является условной в том смысле, что она зависит от набора кластеров $\vec{N}$. Поскольку $E_i$ зависит от $T_{\vec{N}} = T(E, \vec{N})$, наиболее вероятное распределение энергии по кластерам соответствует случаю, когда все кластеры имеют одну температуру.

Относительная величина каждого члена в (6) представляет собой вероятность существования набора кластеров $\vec{N}$

$$P(\vec{a} | \vec{N}) = \prod_1^N G_i(E_i)^{N_i} / N_i! / \Omega(\vec{a}). \qquad (13)$$

В соответствии с этим функция распределения числа кластеров данного размера $i$ определяется выражением

$$f(\vec{a}|N_i) = \sum_{\vec{X}} \delta(N_i - X_i) P(\vec{a} | \vec{X}), \qquad (14)$$

где $\delta(...)$ – символ Кронекера, а функция распределения кластеров по размерам формулой

$$f(\vec{a} | i) = \sum_{N_i} N_i f(\vec{a} | N_i) = \sum_{N_i} \sum_{\vec{X}} N_i \delta(N_i - X_i) P(\vec{a} | \vec{X}). \qquad (15)$$

Первая функция удовлетворяет условию $\sum_i \sum_{N_i} f(\vec{a}|N_i) = 1$, а вторая $\sum_i i f(\vec{a} | i) = N$.

Далее для компактности изложения воспользуемся формализмом условных термодинамических потенциалов [17]. Применительно к рассматриваемой системе он состоит в следующем.

По аналогии с известным соотношением $S(\vec{a}) = k \ln \Omega(\vec{a})$, где $S(\vec{a})$ – энтропия системы, каждому члену в правой части (6) можно ставить в соответствие условную энтропию системы, отвечающую существованию в системе группы кластеров $\vec{N}$:

$$S(\vec{a} | \vec{N}) = k \ln[\prod_1^N G_i(E_i)^{N_i} / N_i!], \qquad (16)$$

тогда (13) записывается в виде

$$P(\vec{a} | \vec{N}) = \exp\{[S(\vec{a} | \vec{N}) - S(\vec{a})]/k\}. \qquad (17)$$

Аналогично можно ввести условную энтропию, соответствующую наличию $N_i$ кластеров $i$-го размера

$$\exp[S(\vec{a} | N_i)/k] = \sum_{\vec{X}} \delta(N_i - X_i) \exp[S(\vec{a}|\vec{X})/k], \qquad (18)$$

так что (14) принимает вид

$$f(\vec{a}\mid N_i)=\exp\{[\,S(\vec{a}\mid N_i)-S(\vec{a})]/k\}, \qquad (19)$$

и условную энтропию, отвечающую наличию кластера размера *i*

$$\exp[S(\vec{a}\mid i)/k]=\sum_{N_i} N_i \exp[S(\vec{a}\mid N_i)/k], \qquad (20)$$

с учетом которой (15) переходит в

$$f(\vec{a}\mid i)=\exp\{[S(\vec{a}\mid i)-S(\vec{a})]/k\}. \qquad (21)$$

Для того чтобы провести вычисления, необходимо выработать процедуру суммирования по $\vec{N}$ и определить вид фазового объёма кластера $G_i(N_i)$.

Суммирование по группам кластеров $\vec{N}$ в (6) осложняется наличием условий (7) и (8). По этой причине обычно предпочитают иметь дело не с микроканоническим, а с большим каноническим ансамблем, в котором нет ограничений на число частиц в системе и на полную энергию системы.

При этом естественным образом вносятся два предположения:
➢ существуют кластеры сколь угодно большого размера
➢ таких кластеров может быть много.

Последнее предположение существенно упрощает задачу и в том плане, что дает возможность использовать для нахождения функции распределения кластеров по размерам метод наиболее вероятного распределения.

В реальной системе размер кластера ограничен числом молекул в системе *N*, а число кластеров размера порядка *N* не может превышать нескольких единиц. Более того, в системе может существовать только один кластер с размером более чем целая часть числа *N/2*. Для ненасыщенного пара (достаточно высокие энергии) такое различие между реальной системой и большим каноническим ансамблем несущественно, так как в этом случае вероятность появления больших кластеров в системе ничтожно мала и температура в системе в основном контролируется числом мономеров и остается практически постоянной.

Однако при переходе к пересыщенному пару (низкие энергии) ситуация радикально изменяется. В этом случае система в основном равновесном состоянии содержит кластеры с размерами порядка *N*, поэтому условие ограниченности системы становится существенным. При этом температура в системе может претерпевать большие изменения, и она главным образом контролируется теми кластерами, появление которых в системе приводит к более сильному снятию пересыщения в системе по сравнению с каноническим ансамблем.

Для упрощения суммирования будем использовать "однокапельное" приближение, введенное в [11, 12]. Суть его заключается в предположении, что в состоянии гетерогенного равновесия система содержит только один большой кластер размера порядка *N*, а остальные частицы в виде мономеров и малых кластеров составляют пар, окружающий большой кластер. Размер большого кластера может флуктуировать.

Число частиц в большом кластере (называемом в дальнейшем каплей) будем обозначать через *g*, а число частиц в малых кластерах – через *i*.

В соответствии с "однокапельным" приближением каждый член в правой части (6) запишем в виде

$$G_g(E_g)\prod_{i=1}^{i_0} G_i(E_i)^{N_i}/N_i!. \qquad (22)$$

Он отвечает пересыщенному пару. Верхний предел произведения $i_0$ определяется условием сохранения числа частиц в системе (7), которое в данном случае принимает вид

$$\sum_{i=1}^{i_0} iN_i + g = N. \qquad (23)$$

Заметим, что в соответствии со смыслом «однокапельного» приближения $i_0 < g$. Для малых кластеров можно принять, что $N_i \gg 1$. Тогда используя формулу Стирлинга $N_i! \approx (N_i/e)^{N_i}$ и применяя к (6) метод неопределенных множителей Лагранжа, то есть варьируя выражение (22) по $N_i$ с учетом условий (7) и (8), мы можем найти наиболее вероятные наборы кластеров $\tilde{N}_i$, которые вносят наибольший вклад в (6). Действуя таким образом, получаем

$$\tilde{N}_i(g) = G_i(E_i)\exp[\beta(\mu_1 i - E_i)], \tag{24}$$

где $\mu_1$ – химический потенциал пара, величина которого зависит от наличия и размера капли $g$.

Чтобы определить $G_i(N_i)$, воспользуемся известными интегральными соотношениями между статистической суммой и статистическим весом [18]

$$Q_{\tilde{N}}(N,V,T) = \int_0^\infty \rho_{\tilde{N}}(\vec{a})\exp(-\beta E)dE = L[\rho_{\tilde{N}}(\vec{a})], \tag{25}$$

$$\Omega_{\tilde{N}}(\vec{a}) = \int_0^E \rho_{\tilde{N}}(\vec{a})dE. \tag{25'}$$

Здесь $\rho_{\tilde{N}}(\vec{a})$ – плотность состояний изолированной системы, $Q_{\tilde{N}}(N,V,T)$ – статистическая сумма системы для данного набора $\vec{N}$, равная $Q_{\tilde{N}}(N,V,T) = \prod_{i=1}^{N} q_i^{N_i}/N_i!$, где $q_i$ определяется формулой

$$q_i = V n_\infty \exp\{-\beta[\mu_{l\infty} i + \alpha(i^{2/3} - 1)]\}, \tag{26}$$

которая получается интегрированием известной формулы $E_i = kT^2 \partial \ln q_i/\partial T$ с учётом (12) [11]. Здесь $\mu_{l\infty}$ – химический потенциал жидкости, $n_\infty = n_\infty(T)$ – плотность насыщенного пара. Соотношение (25) представляет собой преобразование Лапласа $L[\rho_{\tilde{N}}(\vec{a})]$ от плотности состояний. Обращая это преобразование и используя соотношение (25') найдём статистический вес системы для наиболее вероятной группы кластеров $\tilde{\vec{N}}(g)$

$$\Omega_{\tilde{\tilde{N}}}(\vec{a}) = Q_{\tilde{\tilde{N}}}(N,V,T_g)\exp(\beta_g E), \tag{27}$$

где $\beta_g = 1/kT_g$ определяется из закона сохранения энергии

$$\sum_{i=1}^{i_0} \tilde{N}_i E_i + E_g = E. \tag{28}$$

При этом мы полагаем, что поверхностное натяжение $\sigma_{l\infty}$ линейно зависит от температуры. Далее из соотношений (27), (28) имеем

$$G_i(E_i) = q_i(N,V,T_g)\exp(\beta_g E). \tag{29}$$

В результате формула (24) приобретает вид

$$\tilde{N}_i(g) = q_i(N,V,T_g)\exp(\mu_1 i), \tag{30}$$

совпадающий по форме с распределением малых кластеров для канонического ансамбля.

Отличие между ними заключается в том, что теперь температура зависит от наличия и размера капли в системе.

Ограничиваясь учетом только максимальных членов и используя (22) статистический вес (6) можно записать в виде

$$\Omega(\vec{a}) = \sum_{g=2}^{N} G_g \prod_{i=1}^{N} G_i^{\tilde{N}_i}/\tilde{N}_i!. \tag{31}$$

В терминах условных энтропий соответственно имеем

$$\exp[S(\vec{a})/k] = \sum_{g=2}^{N} \exp[S(\vec{a}|\tilde{\vec{N}}(g))/k]. \qquad (32)$$

Придадим выражению (26) более удобный для дальнейшего использования вид. При этом ограничимся случаем, когда пар представляет собой идеальный газ. Воспользуемся известным соотношением

$$\mu_1 = \mu_{l\infty} + kT\ln[(N-g)/Vn_\infty], \qquad (33)$$

которое получается путём интегрирования термодинамического равенства
$d\mu = -sdT + vdp$ вдоль линии $T = const$ при условии $\mu_1 = \mu_{l\infty}$ для $N - g = Vn_\infty$.

Подставляя комплекс $Vn_\infty$ из (33) в (26), получаем

$$q_i = (N-g)\exp\{-\beta_g[\mu_{l\infty}(i-1) + \mu_1 + \alpha(i^{2/3}-1)]\}. \qquad (26')$$

С учётом этого формула (30) приобретает вид

$$\tilde{N}_i = (N-g)\exp\{-\beta_g[(\mu_{l\infty}-\mu_1)(i-1) + \alpha(i^{2/3}-1)]\}. \qquad (34)$$

Выражение для условной энтропии, отвечающей наличию в системе группы кластеров $\tilde{\vec{N}}(g)$, можно найти путём сравнения (31) и (32). Используя формулу Стирлинга для $N_i!$, соотношения (29), (34) и (26') получаем

$$T_g S(\vec{a}|\tilde{\vec{N}}(g)) = E - \mu_{l\infty}g + kT_g \ln(n_\infty V) - \alpha(g^{2/3}-1) - \mu_1(N-g) - kT(N-g), \qquad (35)$$

где $\mu_1$ подчиняется формуле (33) при соответствующих значениях $T_g$ и $g$, указанного в качестве аргумента в левой части (35).

Теперь можно определить функцию распределения кластеров по размерам $f(\vec{a}|i)$ (21). Используя соотношения (18)-(21), и учитывая (34), (35) получаем следующее выражение для условной энтропии, соответствующей $f(\vec{a}|i)$

$$\exp[S(\vec{a}|i)/k] = \sum_{g=2}^{N} \tilde{N}_i \exp[S(\vec{a}|\tilde{\vec{N}}(g))/k] + \exp[S(\vec{a}|\tilde{\vec{N}}(i))/k], \qquad (36)$$

где $\tilde{N}_i$ определяемые формулой (34), являются функциями $g$, указанного в качестве аргумента в условных энтропиях. Сумма в правой части (36) пропорциональна вероятности появления кластера размера $i$ как члена последовательности, отвечающей насыщенному пару над каплей размера $g$, а последний член – как капли размера $i$.

В сумме следует полагать $N_i = 0$ при $i > i_0$, где $i_0$ определяется законом сохранения числа частиц (23), а также при $i > i_*$, где критический размер кластера $i_*$ определяется из условия $\partial \tilde{N}_i / \partial i = 0$. Это условие даёт

$$i_* = \left[\frac{2\alpha}{3kT_g \ln(s_g \cdot (N-g)/N)}\right]^3, \qquad (37)$$

где $s_g = N/Vn_\infty(g)$ – величина пересыщения.

Энтропия системы $S(\vec{a})$ в (21) определяется формулой (32). В случае, когда $S(\vec{a}|\tilde{\vec{N}}(g))$ имеет острый максимум по $g$, соответствующий образованию стабильной капли размера $g_s$, функция распределения $f(\vec{a}|i)$ допускает простое представление "по частям" – в виде функ-ции распределения малых кластеров и функции распределения капель. Это имеет место при больших значениях пересыщения.

Разложим $S\left(\vec{a}|\tilde{\vec{N}}(g)\right)$ вблизи $g_s$ в ряд Тейлора до членов второго порядка включительно

$$S\left(\vec{a}|\tilde{\vec{N}}(g)\right) \approx S\left(\vec{a}|\tilde{\vec{N}}(g_s)\right) + \frac{1}{2}(S'')_{g_s}(g-g_s)^2. \qquad (38)$$

Тогда переходя в (31) от суммы к интегралу (поскольку $g_s \gg 1$) и заменяя пределы интегрирования на $-\infty$ и $+\infty$ (так как максимум предполагается острым), для энтропии системы получаем

$$\exp[S(\vec{a})/k] = \int_{-\infty}^{+\infty} \exp[S\left(\vec{a}|\tilde{\vec{N}}(g)\right)/k]dg = (2\pi\sigma^2)^{1/2} \exp[S\left(\vec{a}|\tilde{\vec{N}}(g_s)\right)/k], \qquad (39)$$

где среднеквадратичное отклонение $\sigma$ числа частиц в стабильной капле получается из

$$\sigma^2 = -k/S''(g_s) = [1/(N-g_s) - 2\alpha/9kT_{g_s}g_s^{4/3} + (q_\infty - 2\alpha/3)(dT_g/dg)_{g_s}/kT_{g_s}^2]^{-1}. \qquad (40)$$

Размер стабильной капли $g_s$ удовлетворяет уравнению Кельвина

$$\ln[(N-g_s)/Vn_\infty] = 2\alpha/3kT_{g_s}g_s^{1/3}, \qquad (41)$$

которое является следствием условия $\partial S\left(\vec{a}|\tilde{\vec{N}}(g)\right)/\partial g = 0$. В последней формуле надо учесть, что концентрация насыщенных паров $n_\infty$ зависит от температуры по закону

$$n_\infty = const \exp(-q_\infty/kT_g). \qquad (41')$$

При фиксированных $N$ и $V$ уравнение (41) определяет зависимость $g_s$ от $E$. Формально (41) имеет два решения: $g_s > g_c$ и $g_s < g_c$, где $g_c$ определяется из условия $S''(g_c) = 0$, которое в соответствии с (40) может быть записано в виде

$$1/(N-g_c) = 2\alpha/9kT_{g_c}g_c^{4/3} - (q_\infty - 2\alpha/3)(dT_g/dg)_{g_c}/kT_{g_c}^2. \qquad (42)$$

Отметим, что данному значению $g_c$ по формуле (42) соответствует значение $E_c'$, которое имеет смысл критической энергии, но не совпадает с действительным значением $E_c$, определяемым из условия $\partial^2 S(\vec{a}|i)/\partial i^2 = 0$, поскольку само уравнение (42) получено в приближении $E < E_c$. Корень $g_s > g_c$ отвечает стабильной капле.

С учётом (38) и (39) функция распределения больших кластеров (капель) по размерам будет иметь вид

$$f(\vec{a}|g) = \exp[-(g-g_s)^2/2\sigma^2]/(2\pi\sigma^2)^{1/2}. \qquad (43)$$

В том же приближении для распределения малых кластеров имеем

$$f(\vec{a}|i) = \int_{-\infty}^{+\infty} \tilde{N}_i(g)f(\vec{a}|g)dg \approx \int_{-\infty}^{+\infty} \tilde{N}_i(g)\delta(g-g_s)dg = \tilde{N}_i(g_s). \qquad (44)$$

Используя (40) и (41) нетрудно перейти к термодинамическому пределу $N \to \infty$, $V \to \infty$, $N/V = const$.

Отметим, что при $T = const$ мы автоматически получаем распределение кластеров по размерам в изотермической системе, полученное в [11, 12]. Сравнение с результатами этих работ показывает, что при одинаковых значениях $g_s$ и пересыщения $s$ отклонения от равновесия (флуктуации) в изолированной системе выражены слабее, чем в изотермической системе, что является следствием наличия дополнительного условия сохранения энергии в изолированной системе.

Однако, эта разница становится несущественной в термодинамическом пределе, поскольку при этом относительные флуктуации в обеих системах становятся пренебрежимо малыми.

## 2. Определение равновесного распределения из уравнения кинетики нуклеации

Сначала рассмотрим изотермическую систему. Основные результаты этой части получены на основе рассмотрения кинетического уравнения нуклеации, описывающего процесс конденсации

$$\frac{\partial n_g}{\partial t} = -\frac{\partial}{\partial g}\{(W_g^+ - W_g^-)n_g - \frac{1}{2}\frac{\partial}{\partial g}[(W_g^+ + W_g^-)n_g]\}, \qquad (45)$$

которое справедливо для $g \gg 1$ [19,20]. Здесь $n_g$ – концентрация кластеров, содержащих $g$ молекул, $W_g^+$ и $W_g^-$ – средние потоки молекул на поверхность кластера $g$-го размера и с поверхности того же кластера. Первый член в правой части (45) описывает направленное движение, а второе – диффузионное движение кластера в пространстве размеров кластера.

Равновесное распределение $n_g^0$ удовлетворяет условию детального баланса

$$\frac{1}{2}\frac{d}{dg}[(W_g^+ + W_g^-)n_g^0] = (W_g^+ - W_g^-)n_g^0, \qquad (46)$$

Это уравнение после интегрирования даёт

$$n_g^0 = C \exp\{2\int_1^g [(W_g^+ - W_g^-)/(W_g^+ + W_g^-)]dg\}/(W_g^+ + W_g^-),$$

где $C$ – постоянная интегрирования.

Для определения скорости конденсации молекул пара на каплю $W_g^+$ будем предполагать, что система состоит из одной единственной капли (большого кластера) размера $g$ и $N-g$ мономеров. Поэтому $W_g^+ = ap(N-g)S_g / N\sqrt{2\pi mkT}$.

Скорость испарения молекул с поверхности капли равна

$$W_g^- = ap_\infty S_g \exp(2\alpha/3kTg^{1/3})/\sqrt{2\pi mkT}.$$

Здесь $a$ – коэффициент конденсации, $S_g$ – площадь поверхности капли, $m$ – масса одной молекулы, $p_\infty$ и $p$ – давление насыщенных паров над плоской поверхностью жидкости и давление в системе в отсутствии капли.

Найдём экстремальные точки функции распределения $n_g^0$. Они определяются из уравнения $dn_g^0/dg = 0$.

Если принять разумное предположение, что $|d\ln W_g^\pm/dg| \ll |d\ln n_g^0/dg|$, то из (46) следует, что точки экстремума удовлетворяют условию $W_g^+ = W_g^-$. Оно эквивалентно уравнению Кельвина

$$\ln\left(S\frac{N-g}{N}\right) = \frac{2\alpha}{3kTg^{1/3}}, \qquad (47)$$

где $S = p/p_\infty$ – пересыщение.

Можно показать, что уравнение (47) для заданных $N$ и $T$ (или $V$ и $T$) имеет два решения при $S > S_c$, одно решение при $S = S_c$, и не имеет ни одного решения при $S < S_c$, где критическое пересыщение $S_c$ определяется из системы уравнений

$$\ln\left(S_c\frac{N-g_c}{N}\right) = \frac{2\alpha}{3kTg_c^{1/3}},$$

$$1/(N - g_c) = 2\alpha/9kTg_c^{4/3}.$$

Второе из этих уравнений выражает тот факт, что при $S = S_c$ оба экстремума функции $n_g^0$ должны слиться (при этом $d^2 n_g^0 / dg^2 |_{g_c} = 0$).

Далее будем рассматривать случай $S > S_c$. При этом решение уравнения (47) с меньшим численным значением ($g_u$) соответствует кластеру критического размера, а другое решение $g_s$ – стабильному кластеру. В соответствии с "однокапельным" приближением наиболее вероятно существование единственной капли, то есть.

$$\int_{i_0}^{N} n_g^0 dg = 1, \quad (48)$$

где $i_0 \sim g_u$.

При $S \gg S_c$ равновесная функция $n_g^0$ имеет острый максимум по $g$, соответствующий образованию стабильной капли размера $g_s$. Тогда $n_g^0$ допускает простое представление "по частям" – в виде функции распределения малых кластеров и функции распределения капель.

Разложим $\ln n_g^0$ в ряд Тейлора вблизи $g = g_s$, ограничиваясь квадратичными членами

$$\ln n_g^0 = \ln n_{g_s}^0 - (g - g_s)^2 / 2\sigma^2,$$

где $\sigma^2 = [1/(N - g_s) - 2\alpha/9kTg_s^{4/3}]^{-1}$ – среднеквадратичное отклонение числа частиц в стабильной капле.

С учётом (48) для равновесного распределения капель по размерам получаем

$$n_g^0 = \exp[(g - g_s)^2 / 2\sigma^2]/\sigma\sqrt{2\pi}. \quad (49)$$

В том же приближении для малых кластеров ($i < i_0$) получаем ($W_1^- = 0$)

$$n_i^0 = (N - g_s)\exp\{2\int_1^i [(W_g^+ - W_g^-)/(W_g^+ + W_g^-)]dg\}W_1^+/(W_i^+ + W_i^-). \quad (50)$$

Равновесное распределение в конечной изолированной системе определяется по вышеприведённой схеме, только здесь надо учесть, что теперь температура в системе зависит от размера капли. Она удовлетворяет уравнению

$$\frac{3}{2}NkT_g = E + q_\infty(g - 1) - \alpha\left(1 - \frac{T_g}{\sigma_{l\infty}}\frac{d\sigma_{l\infty}}{dT_g}\right)(g^{2/3} - 1),$$

где $q_\infty$ – энергия связи на одну молекулу в жидкой фазе.

С учётом вышесказанного равновесное распределение капель и малых кластеров выражаются формулами (49) и (50), при этом дисперсия имеет вид

$$\sigma^2 = [1/(N - g_s) - 2\alpha/9kT_{g_s}g_s^{4/3} + (q_\infty - 2\alpha/3)(dT_g/dg)_{g_s}/kT_{g_s}^2]^{-1},$$

а размер стабильной капли удовлетворяет уравнению $\ln[(N - g_s)/Vn_\infty(g_s)] = 2\alpha/3kT_{g_s}g_s^{1/3}$.

В последней формуле надо учесть, что концентрация насыщенных паров $n_\infty$ зависит от температуры. Однако надо заметить, что в рамках данного подхода эта зависимость не обязательно имеет вид (41').

## 3. Обсуждение теоретических результатов

Данные численного моделирования [2] трехмерной системы, состоящей из конечного числа атомов, взаимодействующих посредством потенциала Леннарда–Джонса, ранее сравнивались с результатами статистическо-термодинамического рассмотрения для изотермической системы в работах [11, 12]. Ввиду хорошего совпадения результатов кинетического подхода с результатами [11, 12], сравнение с данными [2] здесь не приводится.

Поэтому приведём только сравнение полученных результатов с данными для двумерной конечной системы [5, 9, 10].

Аналогично случаю трёхмерной системы, рассмотренному выше, для изолированной двумерной системы средний размер стабильной капли определяется формулой

$$\ln[S(N-g_s/N)] = \alpha/2kT_{g_s} g_s^{1/2}, \qquad (51)$$

а дисперсия

$$\sigma = [1/(N-g_s) - 1/4kT_{g_s} g_s^{3/2} + (q_\infty - \alpha/2)(dT_g/dg)_{g=g_s}/kT_{g_s}^2]^{-1/2}, \qquad (52)$$

где $T_g$ определяется из условия сохранения полной энергии системы

$$E = NkT_g - q_\infty(g-1) + \alpha(g^{1/2}-1). \qquad (53)$$

На рис. 1 дано сравнение функции распределения кластеров по размерам (43), (44) (или (49), (50)) для $q_\infty = 4\varepsilon$, $\alpha = 2\varepsilon$ ($\varepsilon$ – глубина потенциальной ямы) с данными моделирования методом молекулярной динамики изолированной системы частиц, взаимодействующих с помощью потенциала в виде прямоугольной ямы [5], где $E = -20\varepsilon$, $N = 100$.

При этом $\sigma \ll g_s$ и $kT_{g_s} \approx 0.5\varepsilon$, что близка к температуре, полученной в [5]. Причина несохранения полного числа частиц в том, что в формулах (51)-(53) пренебрегли вкладом димеров, тримеров и тому подобное.

Следует заметить, что функция распределении малых кластеров (44) (или (50)) не даёт хорошего количественного согласия с данными машинного моделирования, поскольку в области малых размеров кластеров ( i < 10 ) классическое "капиллярное" приближение плохо работает.

В работе [9] методом Монте-Карло изучалась двумерная решёточная модель газа из *N* частиц в квадратной решётке, состоящей из *K* узлов (модель Изинга) при $T = const$. При $N/K \approx 0.01$ и $kT/\varepsilon_0 = 1/3$ ($\varepsilon_0$ [9] энергия связи между двумя частицами, находящихся в соседних узлах решётки) в равновесном состоянии наблюдался один большой кластер (капля) размера $g_s$, окружённый паром из мономеров и малых кластеров.

Там же приводились значения среднего размера капли и среднего числа мономеров, а также зависимости числа мономеров и размера капли от времени. Функция распределения числа частиц в капле и функция распределения числа мономеров (при тех же значениях *T, K, N* и для той же решётки) были найдены в [10] на основе прямого подсчёта статистических сумм кластеров и было получено хорошее согласие с данными [9].

Следует также заметить, что для двумерной решёточной модели точно известны как кривая существования фаз, так и поверхностное натяжение, тогда как в трёхмерной леннард-джонсовской модели газа не известны достаточно точно ни то и ни другое.

Это является большим преимуществом двумерной решёточной модели газа и одной из причин изучения таких систем.

Легко можно показать, что и для двумерной решёточной модели газа всё изложенное выше остаётся в силе, а именно: средний размер капли и дисперсия вычисляются по формулам (51), (52) (причём, теперь надо полагать $T_g = T = const$, $S = N/Kn_\infty$), а функция распределения числа частиц в капле формулой (49).

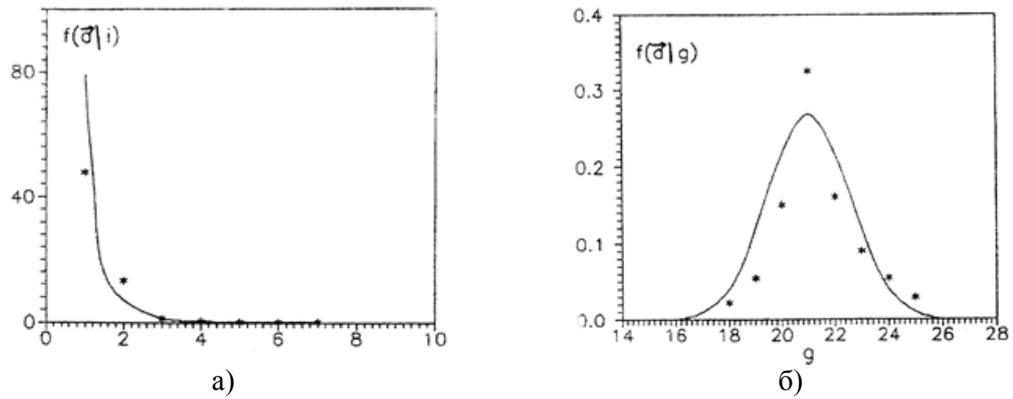

**Рис. 1.** Функция распределения по размерам в конечной двумерной изолированной системе: а) – малых кластеров, б) – капель, * – данные моделирования методом молекулярной динамики [5], сплошная линия – теория (формулы (43) и (44)).

Функция распределения числа мономеров $f(N_1)$ определяется по аналогичной (49) формуле с той же дисперсией и средним значением, равным $N - g_s$ (опять же вкладом димеров, тримеров и тому подобное пренебрегаем).

Для квадратной решётки [21]

$$\sigma_{l\infty} = \varepsilon_0/2 - kT\ln[(1+u)/(1-u)], \qquad (54)$$

$$n_\infty = u^4(1 + 4u^2 + 17u^4 + ...), \qquad (55)$$

$$\alpha = 2\pi^{1/2}\sigma_{l\infty}, \qquad (56)$$

где $u = \exp(\varepsilon_0/2kT)$. При температуре $kT/\varepsilon_0 = 1.3$ значение $\alpha = 1.23$, а величина $n_\infty = 3.11 \cdot 10^{-3}$.

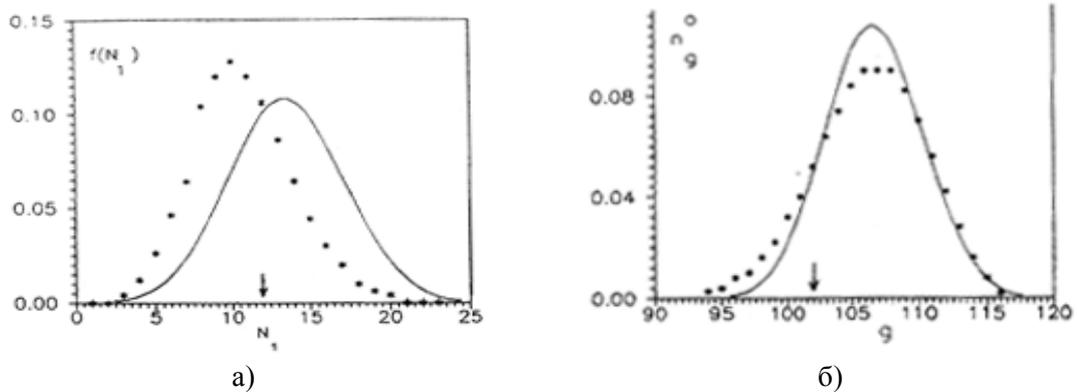

**Рис. 2.** Функция распределения числа мономеров а) и функция распределения капель по размерам б) в двумерной изометрической системе (решёточная модель газа – модель Изинга). * – Данные [10], сплошная линия – теория (формула (49)). Стрелками указаны значения абсцисс, полученные методом Монте-Карло в [9].

На рис. 2, 3 приведено сравнение полученных нами результатов (с учётом (54)-(56)) с данными работ [9, 10]. Рис. 2 соответствует случаю *N = 120, K = 3600*, при этом пересыщение *S = 10.72*, средний размер капли $g_s = 106.6$, дисперсия $\sigma = 3.68$, среднее число мономеров $<N_1> = 13.4$.

А рис. 3 относится к случаю *N = 137, K = 6400*, чему отвечают *S = 6.88*, $g_s = 113.3$, $<N_1> = 23.7$.

На обоих рисунках стрелками показаны значения среднего размера стабильной капли $g_s^{MC}$ и среднего числа мономеров $<N_1>^{MC}$, полученные методом Монте-Карло в [9]: для рис. 2 $g_s^{MC} = 102$, $<N_1>^{MC} = 12$, а для рис. 3 $g_s^{MC} = 106$, $<N_1>^{MC} = 22$.

И здесь видно согласие теории с данными численного моделирования. Из всего сказанного можно сделать вывод, что "капиллярное" приближение может описывать свойства кластера, по крайней мере, при $g > 100$. Полученное в работе распределение мономеров и распределение по размерам капли полностью не совпадают с данными численного моделирования, причиной чего, в частности, могут быть сделанные в работе приближения, приведшие к симметричному распределению по размерам капли.

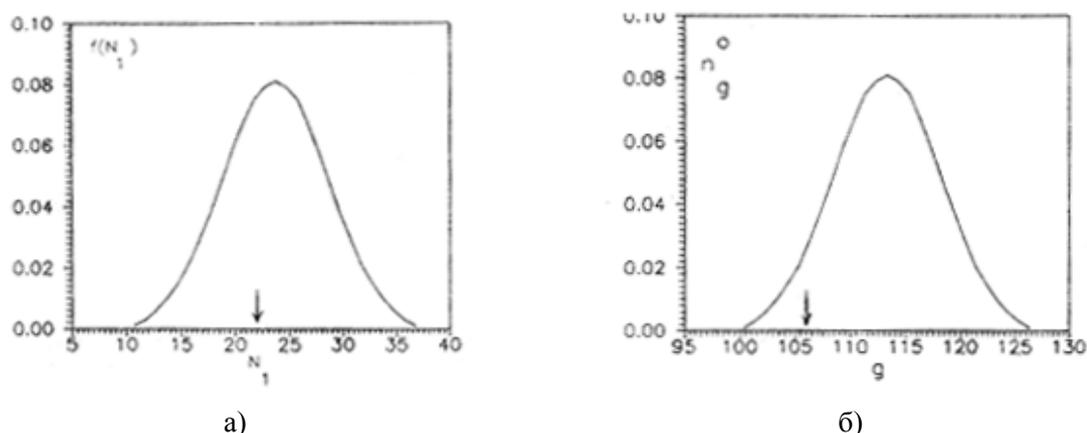

а)                                                                        б)

**Рис. 3.** Функция распределения числа мономеров а) и функция распределения капель по размерам б) в двумерной изометрической системе (решёточная модель газа – модель Изинга). * – данные [10], сплошная линия – теория (формула (49)). Стрелками указаны значения абсцисс, полученные методом Монте-Карло в [9].

Полученное выше хорошее согласие даёт основание полагать, что полученные нами простые приближенные выражения для функции распределения кластеров по размерам будут полезны при оценке равновесного состояния в других аналогичных системах для случая высоких пересыщений.

Для более точного определения функции распределения кластеров по размерам требуется точное знание зависимости фазового объема от энергии и размера кластера. Определение этой зависимости представляет собой чрезвычайно сложную задачу с теоретической, вычислительной и экспериментальной точек зрения (смотрите, например, [22-28]). Знание распределения кластеров по размерам необходимо в разработке нанотехнологий и для практического применения [29-31].

**Выводы**

1. Капиллярное приближение, предполагающее, что кластер состоит из внутренних атомов, имеющих свойства атома в макроскопическом массиве жидкости, и поверхностных атомов, описываемых поверхностным натяжением макросистемы, применимо, по крайней мере, для кластеров, состоящих из ста и более атомов.

2. Однокапельное приближение, предполагающее существование одного кластера (капли) большого размера и кластеров малых размеров, находящихся в равновесии с большим кластером, дает корректные результаты для конечной изолированной системы.

3. Равновесная функция распределения кластеров (наночастиц) по размерам, полученная с использованием однокапельного и капиллярного приближений с помощью методов статистической термодинамики для системы, состоящей из конечного числа молекул, находящихся в конечном замкнутом объеме при постоянной полной энергии количественно описывает данные численного моделирования.

4. Равновесная функция распределения кластеров по размерам, полученная в результате решения кинетического уравнения нуклеации с использованием однокапельного и капиллярного приближений, как для изолированной, так и для изотермической системы, количественно описывает данные компьютерного моделирования конечных систем.

**Литература**

# Equilibrium size distribution function of clusters in finite system

© **Ikhtier Holmamatovich Umirzakov**[+]
*The laboratory of modeling. Kutateladze Institute of Thermophysics of SB of RAS.*
*Prospect Lavrenteva, 1.Novosibirsk 630090 Russia.*
*Tel.: +7 (383) 354-20-17. E-mail: tepliza@academ.org…….*

***Keywords***: *isolated and isothermal systems, microcanonical and canonical ensembles, nanoparticle, cluster, finite system, size distribution function, thermodynamic equilibrium.*

## Abstract

The equilibrium size distribution function of clusters (nanoparticles) in the system of finite number of molecules (atoms) in finite closed volume with constant total energy (isolated system) is found using methods of statistical thermodynamics. The distribution function is found from kinetic equation of nucleation using one-drop approximation for both isothermal and isolated systems. The results obtained are compared with computer simulation data of finite two-dimensional system.